\documentstyle[11pt,aaspp4]{article}
\begin{document}

\title{Did Zhao \& Qin Solve the Apparant Conflict Beween Gravitational Lens Time Delays, Dark Matter
  and the Hubble Constant?}

\author{C. S. Kochanek}
\affil{Harvard-Smithsonian Center for Astrophysics \\
       60 Garden Street \\
       Cambridge, MA 02138 \\
       ckochanek@cfa.harvard.edu}

\begin{abstract}
The solution proposed by Zhao \& Qin to the apparent conflict between gravitational lens
time delays, local estimates of the Hubble constant and current expectations for the 
structure of CDM halos is discussed.  Two essential points emerge.  First, the degeneracy
is exactly the same as the local surface density degeneracy previously discussed in
the literature. Second, the proposed mass distribution is inconsistent with CDM
halo models.  The Hubble constant is raised by making the dark matter far less 
centrally concentrated than predicted for CDM halos, much like the changes
suggested for reconciling CDM halos with the rotation curves of dwarf and LSB
galaxies.   Thus, while galaxies still have dark matter, the Zhao \& Qin solution to the
time delay problem requires fundamental changes in the CDM paradigm for halo structure.  
\end{abstract}

\section{Discussion}

In a series of papers (Kochanek [astro-ph/0204043],  [astro-ph/0205319], [astro-ph/0206006])
we have been exploring the relationship between recently measured gravitational lens time
delays, the expected properties of CDM halos, and local estimates of the Hubble constant.
At least at present, there seems to be a problem because reconciling the time delays
with the local estimates of the Hubble constant seems to require constant $M/L$ distributions
in which the mass traces the light, quite different from our theoretical expectations
and other evidence against constant $M/L$ on the relevant scales.  While no position was
taken on the solution to the problem, one suspects it is a combination of problems in
all three elements.  

Unfortunately, we cannot determine the dark matter distributions of the time delay lenses
given the available model constraints, so it is relatively easy to propose alternate mass
distributions for these lenses.  In particular, we demonstrated in Kochanek (2002, [astro-ph/0205319])
that the time delays of gravitational lenses are controlled by the projected surface density 
of the lens in the annulus between
the images for which the delay is measured.  Zhao \& Qin ([astro-ph/0209191], [astro-ph/0209304])
use this degeneracy to create mass distributions with high Hubble constants and
massive dark matter halos.  In essence, they use a constant M/L model for the lens out
to the radius of the outer lensed image.  Since this model has a low surface density
in the annulus between the images it allows a high Hubble constant.  They then attach
a dark halo with a flat rotation curve onto the mass distribution starting at the 
outer most lensed image.  Essentially by Gauss' law, the halo affects neither the lens
model constraints nor the high Hubble constant estimate permitted by the constant
M/L central regions of the model.

\noindent Two simple points have been lost in generating the models.  

First, it is not a new degeneracy.  It is exactly the same degeneracy we discussed in
Kochanek (2002, [astro-ph/0205319]) and was partially discussed by Saha (2000),
Gorenstein et al. (1988) and Falco et al. (1985).  The value of the Hubble constant
found for a given time delay is (basically) proportional to $1-\langle \kappa \rangle$ where 
$\langle \kappa \rangle $ is the
average surface density in units of the critical surface density for lensing in 
the annulus between the images for which the delay is measured.  The Key Project
value of 70km/s/Mpc requires $0.1 < \langle \kappa \rangle < 0.2$ given the delays in the 4 simplest
time delay lenses.   The observed lens geometry constrains the mass enclosed by
the Einstein ring, for a fixed Hubble constant the time delay constrains the 
surface density near the ring, and you can then attach whatever exterior (monopole)
mass distribution you desire without affecting any observable properties of the lens.
Zhao \& Qin simply produce a particular realization of such a model.

Second, while the Zhao \& Qin mass model has an infinitely massive dark matter halo,
it is not a model consistent with the expectations of CDM.  CDM predicts not only
the existence but also the structure of the dark halos surrounding galaxies. 
The Zhao \& Qin model can be thought of as a model in which the 
central cusp of the dark matter distribution has been destroyed and replaced by
a finite core radius at least as large as the Einstein ring of the lens.  
In Kochanek (2002, [astro-ph/0206006]) we illustrated how CDM halo models are 
related to time delays and the Hubble constant, and demonstrated that the two
can be reconciled provided the mass fraction represented by the visible lens
galaxy is large.  Given the globally small baryonic mass fraction of halos,
standard CDM halo models have difficulty reducing the amount of dark matter in the
central regions enough to allow a large Hubble constant.  The Zhao \& Qin model
solves this problem by giving the dark matter a distribution which avoids the 
central regions of the halo.  It is not, however, a standard CDM halo.

In fact, the Zhao \& Qin model is very similar to the solutions proposed for the
rotation curves of dwarf galaxies (see van den Bosch \& Swaters [astro-ph/0006048]
or McGaugh [astro-ph/0107490] and references therein).  The
rotation curves of some dwarf and low surface brightness galaxies seem to be
inconsistent with the relatively high central dark matter densities implied by
the NFW or Moore profiles, suggesting to some that the cusp must somehow be
converted into a large finite core radius (e.g. the Burkert (1995, ApJL 447 25) profile).  
There is, however, considerable debate about the existence and significance of
the conflict between CDM halo models and dwarf/LSB rotation curves.
The Zhao \& Qin model represents the same class of solution -- keep a massive dark matter
halo but greatly reduce its central concentration compared to the cuspy CDM
halo models.  Just as in the dwarf/LSB galaxies, this allows the central 
mass distribution to follow the stellar distribution, and constant M/L 
mass distributions for the time delay lenses lead to values of the Hubble 
constant large enough to agree with local estimates.

While it is certainly correct that the present data on the time delay lenses 
cannot exclude the Zhao \& Qin mass distribution, it is probably ruled out as
a general distribution from other observations.  In particular, where lenses
do constrain the shape of the rotation curve directly, they almost always
favor an essentially flat rotation curve inside the Einstein ring rather than
the more Keplerian form of the Zhao \& Qin model.  The most interesting recent
demonstrations of this have come from Koopmans \& Treu ([astro-ph/0202342],
[astro-ph/0205281]) who combine lensing constraints with measurement of the
velocity dispersions of the lens galaxy to show that the two lenses 
MG2016+112 and 0047-281 have mass distributions corresponding to flat 
rotation curves.  These mass distributions imply $\langle \kappa \rangle \sim 0.5$ on the 
corresponding scales of the time delay lenses and are consistent with $\rho \sim r^{-1}$
central cusps in the dark matter.  While it is
always possible that some time delay lenses have peculiar halo properties,
it seems unlikely that all 4 of the (apparently) simple delay lenses have 
peculiar halo properties.

In summary, Zhao \& Qin are using the known degeneracy between surface density
and gravitational lens time delays to invent a dark matter distribution which
allows for a high Hubble constant.  The solution is to change the structure 
of CDM halos so they are significantly less centrally concentrated (cuspy) than 
present theories.  While this may well be an element of the solution to the problem, 
it is important to recognize the implications of the proposal -- galaxies can
still have dark matter halos, but CDM theory has failed to accurately predict 
their structure.

\end{document}